# Atomic-Scale Patterning of Arsenic in Silicon by Scanning Tunneling Microscopy

*Taylor J. Z. Stock[1]\*, Oliver Warschkow[2], Procopios C. Constantinou[1], Juerong Li[3], Sarah Fearn[1,4], Eleanor Crane[1], Emily V. S. Hofmann[1,5], Alexander Kölker[1], David R. McKenzie[2], Steven R. Schofield[1,6], Neil J. Curson[1,7]\**

1. London Centre for Nanotechnology, University College London, London WC1H 0AH, UK

2. Centre for Quantum Computation and Communication Technology, School of Physics, The University of Sydney, Sydney, NSW 2006, Australia

3. Advanced Technology Institute, University of Surrey, Guildford GU2 7XH, United Kingdom

4. Department of Materials, Imperial College of London, London SW7 2AZ, UK

5. IHP – Leibniz-Institut für innovative Mikroelektronik, Frankfurt (Oder) 15236, Germany

6. Department of Physics and Astronomy, University College London, London WC1E 6BT, UK

7. Department of Electronic and Electrical Engineering, University College London, London WC1E 7JE, UK.

AUTHOR INFORMATION

**Corresponding Authors**
\*Taylor J.Z. Stock t.stock@ucl.ac.uk

\*Neil J. Curson n.curson@ucl.ac.uk

ABSTRACT

Over the last two decades, prototype devices for future classical and quantum computing technologies have been fabricated, by using scanning tunneling microscopy and hydrogen resist lithography to position phosphorus atoms in silicon with atomic-scale precision. Despite these successes, phosphine remains the only donor precursor molecule to have been demonstrated as compatible with the hydrogen resist lithography technique. The potential benefits of atomic-scale placement of alternative dopant species have, until now, remained unexplored. In this work, we demonstrate the successful fabrication of atomic-scale structures of arsenic-in-silicon. Using a scanning tunneling microscope tip, we pattern a monolayer hydrogen mask to selectively place arsenic atoms on the Si(001) surface using arsine as the precursor molecule. We fully elucidate the surface chemistry and reaction pathways of arsine on Si(001), revealing significant differences to phosphine. We explain how these differences result in enhanced surface immobilization and in-plane confinement of arsenic compared to phosphorus, and a dose-rate independent arsenic saturation density of 0.24±0.04 monolayers. We demonstrate the successful encapsulation of arsenic delta-layers using silicon molecular beam epitaxy, and find electrical characteristics that are competitive with equivalent structures fabricated with phosphorus. Arsenic delta-layers are also found to offer improvement in out-of-plane confinement compared to similarly prepared phosphorus layers, while still retaining >80% carrier activation and sheet resistances of <2 kΩ/□. These excellent characteristics of arsenic represent opportunities to enhance existing capabilities of atomic-scale fabrication of dopant structures in silicon, and are particularly important for three-dimensional devices, where vertical control of the position of device components is critical.

**TOC GRAPHICS**

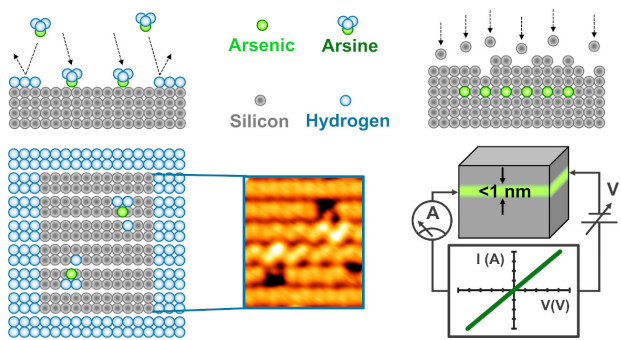

# INTRODUCTION

Challenges associated with the continued miniaturization of electronic devices toward the atomic-scale limit,[1] and proposals for novel atomic-scale device architectures, e.g., to enable quantum information processing,[2] have stimulated renewed interest in the investigation of dopant species in silicon. Of particular interest are methods for controlling the placement of dopant atoms in silicon with atomic-scale precision,[3,4] and understanding the electronic[5,6,7,8,9] and optical[10,11,12] properties of single-atom and few-atom clusters of dopants. The atomic-scale control of dopant atoms in silicon using scanning tunneling microscopy (STM) has to date focused on phosphorus donors, which can be introduced into the silicon matrix using the precursor gas, phosphine ($PH_3$). The chemical reaction of phosphine with the silicon (001) surface has been extensively researched and is now understood in exquisite detail.[13] This chemical process can be spatially controlled at the atomic-scale via the use of STM hydrogen-desorption lithography [14] to produce atomically-defined patterns of phosphorus donors in silicon.[15] This has led to numerous exciting advances in our understanding of silicon-based electronic devices at the atomic-scale,[16,17,18,19] and is a promising avenue for the use of individual donor spins as qubits in a solid state quantum computer. [20,21,22]

Arsenic donors in silicon are particularly interesting, since they have a lower diffusivity and a higher solid solubility in bulk silicon than phosphorus,[23] as well as a higher ionization energy in silicon (53.76 meV, compared to 45.59 meV for phosphorus),[24] a larger atomic radius ($r_{As}$ =115 pm, $r_P$ = 100 pm),[25] larger atomic spin-orbit interaction ($Z_{As}$ = 33, $Z_P$ = 15), and a higher nuclear spin value ($I_{As}$ = 3/2, $I_P$ = ½) than phosphorus. These differing properties present opportunities for atomic-scale device designs with advanced functionality, such as nuclear spin qudits (where a qudit is a generalized quantum information object with n > 2 quantum states),[26,27] and silicon photonic crystal cavity based quantum computation schemes.[28] Furthermore, the ability to position multiple dopant species in silicon with atomic-scale precision would allow independent addressing of each donor species, and will enable new principles of device operation, such as implementations of quantum error correction codes,[29] and optically driven silicon-based quantum gates.[30] Thus, the ability to control the reaction of arsine ($AsH_3$) with Si(001) using STM hydrogen-desorption lithography, in an analogous manner to phosphine, presents enormously exciting opportunities for atomic-scale electronics.

Kipp et al.[31] have studied the adsorption of arsine on Si(001) using STM and X-ray photoemission spectroscopy and suggest that $AsH_3$, adsorbed at room temperature, decomposes into AsH + 2H and saturates at an As concentration of ~20% of a silicon monolayer (ML). Several other reports on the surface chemistry and growth kinetics of arsenic layers on Si(001) from a gas phase arsine precursor[32,33,34,35,36] have been primarily concerned with rapid growth rates and elevated growth temperatures (> 600 °C), which are not compatible with the thermal requirements of STM hydrogen-desorption lithography.[37,38] In general however, these reports have established that arsine gas phase dosing of Si(001) is self-limiting at an arsenic coverage of ~0.25-1 ML, dependent on the growth temperature, and therefore on the degree of $H_2$ desorption. Previous density functional theory (DFT) studies have considered likely dissociation configurations of $AsH_3$,[39] kinetic activation barriers for the removal of the first H atom,[40] and comparative thermodynamic stabilities of dissociated $AsH_3$ and $PH_3$.[41]

Here, we present a combined atomic-scale STM and DFT investigation of the adsorption and thermal decomposition of arsine on Si(001). We establish the full chemical pathway of arsine decomposition and incorporation of arsenic into the Si(001) surface, and also the spatial control of this reaction via STM hydrogen-desorption lithography. In particular, we demonstrate the following: 1) Individual $AsH_3$ molecules rapidly and fully dissociate on Si(001) at room temperature. 2) Arsenic atoms from dissociated $AsH_3$ molecules incorporate substitutionally into the silicon lattice at temperatures below 350 °C, and at all surface coverages. 3) Arsine adsorption on Si(001), at room temperature, is self-terminating at a saturation coverage of 0.24 ± 0.04 molecules/surface silicon atom, independent of gas exposure rate, providing an arsenic coverage well above the metal-insulator transition, allowing for STM fabrication of metallic interconnects. 4) $AsH_3$ molecules selectively adsorb and dissociate in a STM depassivated region of the Si(001)-H surface, while leaving the surrounding H-resist intact; subsequent annealing incorporates the arsenic atoms into the top layer of the silicon surface, without disturbing the H-resist. 5) Arsenic delta-doped layers can be grown by confining the arsenic incorporated surface below an epitaxial silicon capping layer, grown via temperature-programmed, low temperature epitaxy. 6) Ohmic contacts can be made to the buried, and electrically activated, arsenic-doped layer by aluminum deposition. Fulfillment of these six criteria satisfies the necessary requirements for successful STM fabrication of atomic-scale, electrically contacted, arsenic in silicon electronic devices. While we are aware of unpublished progress towards STM patterning of boron in silicon,[42] the present work represents the first development of the capability to atomically pattern any dopant species, other than phosphorus, in silicon, using hydrogen-desorption lithography. Moreover, this work represents a first step in the development of a materials toolbox, where atomic-scale structures can be constructed using multiple dopant elements; an important step closer towards the emerging field of atomic-scale fabrication and manufacturing.

RESULTS AND DISCUSSION

**1. Dissociation of Isolated $AsH_3$ Molecules**

To ascertain the atomic-scale behavior of the $AsH_3$/Si(001) adsorption system, we exposed atomically clean Si(001)2×1 to a low dose of arsine (0.015 L at $5\times10^{11}$ mbar) at room temperature. An example of the surface observed by STM, immediately after dosing, is shown in the filled and empty-state STM images of **Figure 1**a and 1b. Of the features observed, several are readily identified as native Si(001) surface defects,[43] such as dimer vacancies (labeled DV in **Figure 1**a and 1b ) and C-defects (labeled Cd).[44] The new features in these images are attributed to the products of $AsH_3$ dissociation reacting with the surface. These are labelled "type-1" and "type-2". The type-1 features are by far the most common, accounting for more than 75% of all $AsH_3$-associated features detected at these coverages, and these are the principal focus here. The minority type-2 features arise due to steric effects, which we briefly discuss below.

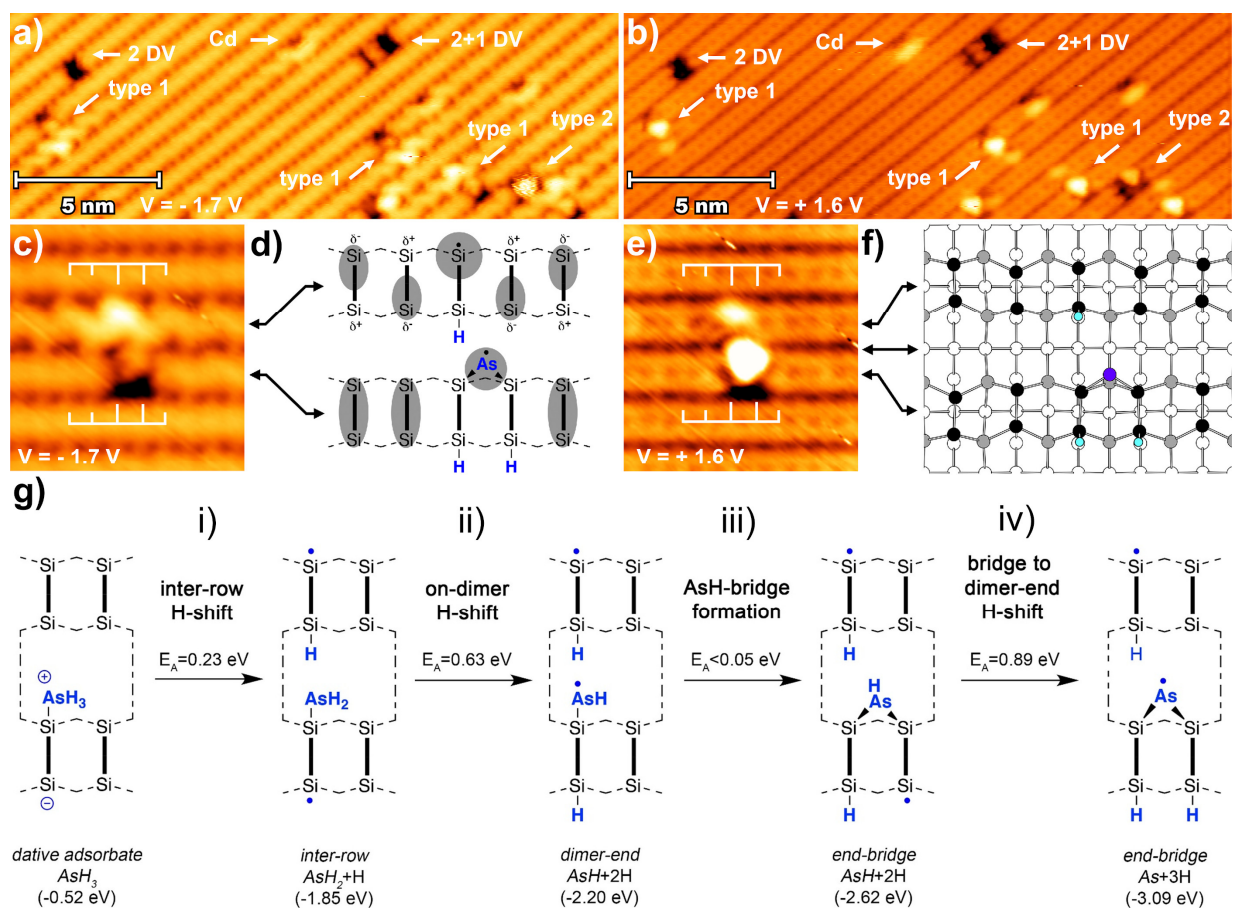

*Figure 1: Adsorption and dissociation of single AsH₃ molecules on Si(001)2×1 a,b) Filled and empty state STM images of low-coverage AsH₃ on Si(001). AsH₃ features are labelled type-1 and type-2, native surface defects are dimer vacancies (DV), double dimer vacancy (2DV), 2+1 dimer vacancy complex (2+1DV), and C-defect (Cd). V = -1.7 V, +1.6 V, I = 0.10 nA. c) Filled-state STM image of a type-1 feature, reticle highlights alignment with Si dimers. V = -1.7 V, I = 0.10 nA. d) Schematic valence diagram details the inter-row end-bridge As+3H structure of the type-1 feature. Grey shading indicates regions imaging bright in STM. e) Empty-state STM image of a type-1 feature, reticles highlight alignment with Si dimers. V = +1.6 V, I = 0.10 nA. f) Structural schematic of inter-row end-bridge As+3H structure showing the positions of first and second layer silicon atoms (black and grey, respectively), and the single As (dark blue) and three H atoms (light blue) provided by the AsH₃ molecule. (g) Five-step reaction path from molecularly-adsorbed AsH₃ to the fully dissociated inter-row end-bridge As+3H structure. Calculated adsorption energies are provided in brackets for each sequential structure. Calculated activation energies, $E_A$, for each step are given above the reaction arrows. The STM images are skewed to account for thermal drift and rotated ~45° from the scan direction such that the dimer-rows align with the horizontal.*

High-resolution, filled- and empty-state images of the type-1 feature are shown in **Figure 1**c and 1e, respectively. The type-1 feature is characterized by two bright protrusions on two adjacent dimer rows. These protrusions appear in both filled and empty states, but differ in their relative brightness. Across the dimer rows, i.e., along the vertical axis of the images in **Figure 1**c and 1e, both protrusions are shifted up from the dimer centers, towards an upper dimer end. Divisions of the overlaid white reticles on the STM images indicate the dimer center positions in the horizontal direction. By reference to these reticles, we see that the upper row protrusion is positioned on top of a single dimer end, while the lower-row protrusion is between two dimer ends. The upper protrusion induces strong dimer pinning, which is apparent in the filled state image (**Figure 1**c) as the characteristic zig-zag appearance along the row.

We assign the type-1 STM feature to the structure shown in **Figure 1**d, which we call the *inter-row end-bridge As+3H* configuration. Grey highlighting in the schematic indicates sites that appear bright in the corresponding filled-state image. These bright sites correspond to the bare Si-Si dimers that surround the adsorbate, and to the two prominent protrusions of the type-1 feature. The two protrusions in the STM images are assigned to the dangling bond end of a Si-Si-H hemihydride dimer on the upper row, and a dimer end-bridge arsenic atom on the lower row (where the remaining two hydrogen atoms are attached to the lower row, directly across from the end-bridge arsenic atom). These two sites are both characterized by a single unpaired electron, i.e. there are half-filled molecular orbitals at these sites that would nominally image brightly in both filled and empty state. The two dimer ends opposite the bridging arsenic atom appear dark in both the filled and empty state image, which is consistent with the electronically saturated, hydrogen-termination at these sites.

**Figure 1**g shows the proposed reaction pathway giving rise to the inter-row end-bridge As+3H configuration observed as the type-1 feature. This kinetically preferred pathway is based on DFT calculations of adsorption energies of the various intermediate structures involved and the activation energies for each step along the path. Each successive step leads to a structure of lower energy, providing the thermodynamic driving force towards the final formation of the inter-row end-bridge As+3H configuration. The first reaction step, labeled (i) in **Figure 1**g, is the dissociation of a datively-adsorbed $AsH_3$ molecule into an $AsH_2$+H species by means of an inter-row hydrogen shift. This reaction transfers one hydrogen atom to the adjacent dimer row, producing a hemihydride dimer. The calculated activation barrier for this reaction step is only 0.23 eV, which corresponds to a reaction that is completed on a timescale of nanoseconds at room temperature. We note that the inter-row hydrogen shift is kinetically preferable to the alternative, (and perhaps more intuitive) on-dimer and inter-dimer hydrogen shifts (not shown in **Figure 1**g), which have calculated activation energies of 0.53 eV and 0.29 eV, respectively. The fact that the inter-row H-shift is preferred, correlates well with the fact that the type-1 STM feature, contains a hemihydride on an adjacent row, and that a majority of adsorbed $AsH_3$ molecules dissociate by this route.

The second reaction step, labeled ii) in **Figure 1**g, is also a hydrogen-shift reaction, which preferentially occurs in the on-dimer direction to produce a structure that we refer to as the dimer-end AsH+2H structure. The calculated activation energy for this reaction is 0.63 eV, corresponding to a reaction timescale of milliseconds at room temperature. The alternative reaction, in which the hydrogen atom shifts in the inter-dimer direction (not shown in **Figure 1**g), has a slightly larger activation barrier of 0.72 eV, and is thus kinetically disfavored. The dimer-end AsH+2H configuration containing a single-valent AsH fragment is a highly transient intermediate corresponding to an extremely shallow minimum on the potential energy surface. This species near-instantly stabilizes into the end-bridge AsH+2H species, in which the AsH fragment bridges between two dimer-ends, labeled iii) in **Figure 1**g.

The final reaction step, labeled iv) in **Figure 1**g, transfers a hydrogen atom from the AsH fragment to the silicon dangling bond at the opposite dimer end. This reaction produces the inter-row end-bridge As+3H structure that corresponds to the type-1 STM feature. The reaction barrier for this final hydrogen-shift is calculated to be 0.89 eV; this is the rate-determining activation energy along the full dissociation path. This barrier is still low enough such that full dissociation of an adsorbing $AsH_3$ molecule into inter-row end-bridge As+3H will be completed on a timescale of tens of seconds. Overall, the calculated reaction pathway is thus highly consistent with observation of the type-1 STM feature as the majority species following $AsH_3$ chemisorption on Si(001).

Despite the kinetic preference for this dissociation pathway, STM images also reveal small numbers of a few alternative dissociation structures, e.g., the type-2 feature seen in **Figure 1**a,b. These arrangements, which are kinetically disfavored on the bare Si(001) surface, are thought to result from steric hindrance provided by neighboring $AsH_3$ adsorbates, or native Si(001) defects. A single example of the type-2 feature is shown in **Figure 1**, in the bottom right of panels a and b, and notably occurs immediately adjacent a type-1 feature. At this low coverage, the type-2 feature comprises roughly 10-15% of the total $AsH_3$ associated STM features, and corresponds to the intra-row end bridge structure, known to be the preferred dissociation configuration for $PH_3$/Si(001).[13]

It is instructive here to briefly compare the dissociation of $AsH_3$ on silicon to the analogous $PH_3$/Si(001) system. The primary $PH_3$ dissociation process at low coverage is a progressive sequence of three principal species, namely, $PH_2$+H, PH+2H, and P+3H.[45] Transitions along this sequence occur on a timescale of minutes, making these transitions directly observable in room-temperature STM experiments (see e.g. Figure. 6 in [45]). In contrast, full $AsH_3$ dissociation into the As+3H species is complete before the surface can be imaged by STM after arsine exposure, leaving the type-1 STM feature as the single dominant species at low coverage. Comparison of our DFT calculations for $AsH_3$/Si(001) with those reported for phosphine[13] provides some insights into how these differences arise. First, hydrogen-shift reactions in the arsine pathway are more exothermic and have lower activation energies than the corresponding reaction in the phosphine pathway. This results in a much faster rate of dissociation for $AsH_3$. Second, a characteristic of the phosphine pathway is that hydrogen-shift dissociation and $PH_2$-fragment diffusion are in close competition; this becomes manifest in the great diversity of $PH_3$ dissociation species formed,[41] and in the observed diffusion of the P atom away from the initial adsorption site.[13] This competition no longer exists for arsine due to the much reduced hydrogen-shift activation barrier. In effect, an adsorbing $AsH_3$ molecule rapidly falls apart where it lands, undergoing minimal rearrangement of the fragments. The result is that the arsenic atom does not have opportunity to diffuse away from the initial landing site and a single structure, namely the *inter-row end-bridge As+3H* structure, is formed as the majority species.

**2. Substitutional Incorporation of Arsenic Atoms into the Silicon Lattice**

Following molecular dissociation of $AsH_3$ on the Si(001) surface, the next step critical to the STM lithography fabrication process, is incorporation of the arsenic atoms into the silicon lattice. **Figure 2**a and **2b** show a clean Si(001) surface exposed to a low dose (0.015 L) of arsine at room temperature (**Figure 2**a), and then subsequently annealed to promote the incorporation of the arsenic atoms into the surface (**Figure 2**b). Before annealing, the surface exhibits many As+3H features (type-1), two of which have been highlighted by arrows. High-resolution filled- and empty-state images of the same surface (inset to **Figure 2**a) provide a close view of a type-1 feature (arrow) and a nearby C-defect. **Figure 2**b shows the same surface (but a different area) after a one-minute thermal anneal at 500 °C. Arrows in **Figure 2**b highlight examples that, by their appearance and by analogy to the behavior of $PH_3$ on Si(001),[46] we assign to Si-As heterodimers formed during the anneal. High-resolution filled and empty state images of two such As-Si heterodimers (inset of **Figure 2**b) reveal an appearance that is very similar to that of P-Si heterodimers, confirming that arsenic incorporation into the surface has taken place.

At the temperature of 500 °C used in the annealing experiment above, all hydrogen atoms are expected to desorb from the silicon surface, and thus **Figure 2**b shows a surface completely depleted of hydrogen. At this temperature, the substituted silicon atoms, which are ejected from the surface as the As-Si dimers are formed, rapidly diffuse away from the ejection site and are eventually captured at the existing step edges, or newly formed ad-dimer rows.[47] This process of controllably substituting arsenic atoms into the silicon lattice is a requirement for compatibility of arsine and arsenic with STM hydrogen resist lithography, as substitution of isolated dopant atoms is necessary if the dopants are to be activated as electron donors in STM fabricated electronic devices.

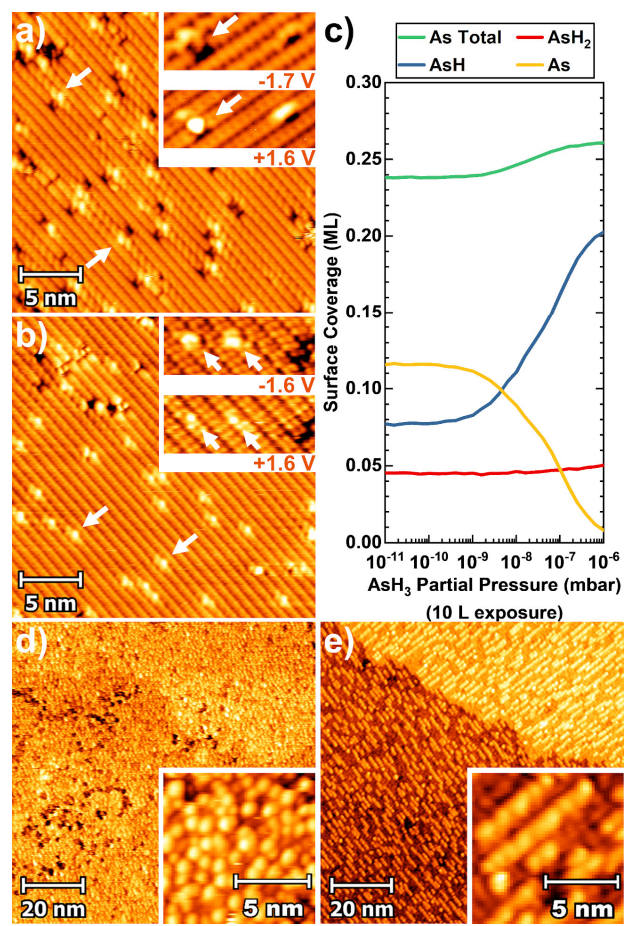

*Figure 2: Substitutional incorporation of isolated arsenic atoms into the silicon lattice and saturation arsine coverage (a) STM image of low coverage (0.02 adsorbates/nm2), room temperature $AsH_3$ on Si(001). Arrows indicate examples of isolated type-1 $AsH_3$ features. V = -1.7 V, I = 0.02 nA The inset shows filled and empty state images of the type-1 feature alongside a C-defect for comparison. (b) Surface from image (a) subsequent to a 500 °C × 1 min anneal. Isolated As-Si heterodimers (HD) are observed in the first layer of the silicon surface, examples are indicated by white arrows. V = -1.6 V, I = 0.10 nA. Two examples of the As-Si HD are shown in filled and empty states in the inset. (c) Kinetic Monte Carlo simulation of total As surface coverage and $AsH_x$ species distribution, as a function of arsine partial pressure. (d) STM image of a Si(001) surface following a 1.5 L (5 × 10-9 mbar × 5 minutes) room temperature, arsine exposure. Inset shows a magnified image of the same surface. V = -2.0 V, I = 0.04 nA. (e) Surface from image (a) subsequent to a 350 °C × 1 min anneal. Inset shows ejected silicon dimer chains indicative of arsenic incorporation. V = -2.0 V, I = 0.05 nA.*

## 3. Saturation Arsine Coverage

Having resolved the elementary AsH$_3$ single-molecule processes and arsenic surface incorporation in the low-coverage regime, we now move on to the saturation-coverage regime. High-density regions of arsenic dopants are important for STM lithography device fabrication, as these conductive 2D areas serve as device interconnects and contacts. **Figure 2**d shows a large-scale STM image of the Si(001) surface dosed to saturation by 1.5 L of arsine. The image shows a disordered pattern of protrusions and depressions as expected for a surface densely covered with the products of AsH$_3$ dissociation.

Hydrogen atoms bonded to Si(001) are immobile at room temperature, and in the stable configuration of the inter-row end-bridge As+3H structure, the arsenic atom is locked into place by the three hydrogen atoms surrounding it. Activation barrier considerations reveal that arsenic incorporation can only take place when the thermal conditions are such that hydrogen atoms can diffuse away from the arsenic atom. The result of annealing the arsine saturated surface to 350°C for 5 minutes is shown in the STM images in **Figure 2**e, this anneal allows the adsorbed hydrogen atoms formed in the course of dissociation to diffuse over the surface, creating space for arsenic atoms to incorporate into the surface dimer layer. That incorporation has occurred is evidenced in **Figure 2**e by the bright linear protrusions running perpendicular to the underlying dimer rows (also see inset to **Figure 2**b). These bright linear protrusions correspond to short ad-dimer chains, formed from the silicon atoms ejected during arsenic incorporation. These ad-dimer chains are a familiar sight from phosphorus incorporation experiments in the context of PH$_3$ lithographic fabrication.[3, 47] Because there is one ejected silicon atom for every incorporated arsenic atom, one can interpret the density of ejected silicon atoms observed in STM images as a direct measure of the density of incorporated arsenic. Using this approach we estimate the incorporated arsenic coverage in samples prepared by a saturation arsine dose of 1.5 L and thermal anneal of 350 °C to be 0.24+/-0.04 ML.

We find that this incorporated arsenic coverage is independent of the arsine saturation dose rate, which is interesting since this is not the case for phosphine (discussed further below). Saturation dosing at different rates can be achieved by changing the arsine partial pressure, with simultaneous adjustment of the total exposure time. When the arsine dose rate is varied by decreasing the arsine partial pressure from $5 \times 10^{-9}$ mbar to $5 \times 10^{-10}$ mbar (along with a dose equalizing change in total exposure time from 5 minutes to 50 minutes), the resulting total arsenic coverage is found to be constant within measurement error. This dose-rate independent saturation arsenic coverage can be explained as follows. (i) At low coverages, AsH$_3$ moieties dissociate rapidly and completely at random locations. (ii) As the density of surface adsorbates increases, the occupied sites begin to sterically hinder the complete dissociation of further incoming AsH$_3$, resulting in the subsequent adsorption of a distribution of partially dissociated AsH$_x$ moieties – a distribution determined by site availability, *not* dissociation rate. (iii) Eventually all adsorption sites are blocked and coverage becomes saturated, leaving a final surface comprised of all possible AsH$_x$ moieties.

The constant incorporated arsenic coverage value, measured across a range of arsine dose rates, is also found to be consistent with predictions from simple kinetic Monte Carlo (KMC) simulations. The results of these simulations are plotted in **Figure 2**c. Here, we use KMC calculations to determine the total arsenic coverage at saturation, for a range of arsine partial pressures, spanning 5 orders of pressure magnitude, in which the distribution of $AsH_x$ fragments is allowed to vary. We assume that isolated molecules will dissociate in the type-1 structure, with a 2.5 dimer footprint discussed in section 1, but that as the surface fills in, and silicon site availability becomes reduced, $AsH_3$ dissociation will be halted at different stages leaving behind AsH and $AsH_2$ fragments. Although the ratio of As, AsH, and $AsH_2$ varies as a function of partial pressure, the total number of arsenic atoms present on the surface remains in the 0.24-0.26 ML range, and the arsenic surface concentration only increases by 2-3% as the dose rate is increased across the full range.

This dose rate independence of arsenic coverage in the saturated $AsH_3/Si(001)$ system is in direct contrast to the $PH_3/Si(001)$ system, where phosphorous coverage at saturation is found to be dose-rate *dependent* at room temperature.[48,49,50] When Si(001) is saturated with $PH_3$ there is competition over available bare surface dimers between the adsorbing $PH_3$ molecules and the dissociating $PH_2$+H species. For $AsH_3$, this competition no longer exists, due to a dissociation rate that is three orders of magnitude more rapid than that of $PH_3$. In effect, any adsorbing $AsH_3$ molecule breaks apart on the Si(001) surface to a degree limited only by the availability of bare Si dimer sites surrounding the molecule.

The much more rapid dissociation of $AsH_3$ also manifests in a lack of ordering in this system. The saturated surface and magnified inset shown in **Figure 2d** show very little evidence of long- or short-range ordering. This is once again in direct contrast to the $PH_3/Si(001)$ system, where $PH_2$ moieties are found to assemble in small patches of local $p(2\times2)$ ordering, with PH moieties aligned along Si dimer rows.[48] In the case of $PH_3$, dissociation and removal of successive H atoms from the $PH_x$ moieties occurs on the time scale of minutes, and the moieties survive long enough to undergo diffusion from the initial adsorption site to energetically preferred, ordered sites. In the $AsH_3/Si(001)$ system, the H atoms are stripped from the molecule much more rapidly, with isolated molecules completely dissociating on the timescale of tens of seconds. The $AsH_x$ moieties do not have opportunity to diffuse before fully dissociating, and at saturation coverage they remain immobilized at the initial, random landing site, producing the completely disordered surface shown in **Figure 2a**.

## 4. Adsorption and Incorporation of Arsenic through a Hydrogen Resist

We now demonstrate that the arsine adsorption, dissociation, saturation, and incorporation on the clean Si(001) surface, as described above, is compatible with hydrogen desorption lithography. For compatibility with the STM fabrication process, $AsH_3$ molecules must selectively adsorb *only* on depassivated Si(001) regions and not on the passivated Si(001):H surface. Furthermore, to allow for maximum in-plane confinement, and fabrication flexibility, arsenic should also incorporate into the silicon lattice at anneal temperatures below the onset of $H_2$ desorption (380 °C). Fulfilment of both these requirements is demonstrated in **Figure 3**. Each panel in this figure corresponds to one step in the STM lithography patterned doping process, with the corresponding steps indicated by the inset schematics on each STM image.

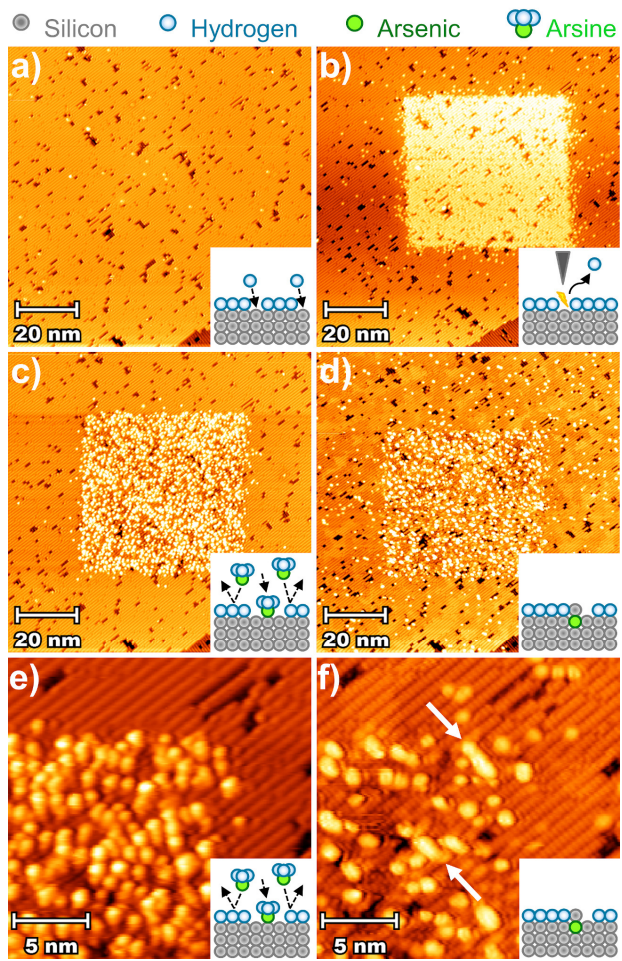

*Figure 3: Compatibility of AsH₃ adsorption, dissociation, and incorporation with STM hydrogen resist lithography* Sequential STM images of the same 100×100 nm area on a Si(001)-H surface showing that the H-resist remains intact and free from adsorbates throughout each step of the STM lithography process. (a) clean hydrogen terminated Si(001) surface prior to STM lithography or arsine exposure. b) selected area hydrogen desorption of a 50×50 nm square (V = +7.0 V, I = 1 nA, s = 100 nm/s) c) 1.5 L arsine dose results in selective adsorption only within the lithographically defined area. d) 350 C × 1 minute anneal, results in incorporation of the adsorbed arsenic atoms as evidenced by presence of ejected Si. e) shows an enlargement of the upper right-hand corner of the patterned square in figure (c). The disordered saturation arsenic layer is clearly contained within only the patterned region, and there is no adsorption on the resist. f) shows an enlargement of the upper right-hand corner of the patterned square in figure (d). Short Si dimer chains, evidence of arsenic incorporation, are indicated by the arrows. The surrounding hydrogen resist remains essentially intact with only a small density of additional dangling bonds produced by the incorporation anneal. V = -2.0 V, I = 0.10 nA.

In the first step, a hydrogen passivated Si(001) surface is prepared, as shown in **Figure 3**a. Each silicon atom has one or two hydrogen atoms bonded to it, except in a few cases where the hydrogen atom is missing, leaving a dangling bond (DB). The DBs have the appearance of bright dots in the image and the DB density is < 1%. Selective removal of a region of the hydrogen 'resist' layer is achieved by scanning over a 50×50 nm region of the surface using an increased bias/current of +7.0 V and 1.0 nA, at a tip raster speed of 100 nm/s (**Figure 3**b). The patterned region, in the center of the image, shows the density of bright dots now approaches 100% coverage of the surface, indicating that most of the hydrogen atoms have been desorbed from the surface, leaving behind the bare silicon.

Dosing the patterned surface of **Figure 3**b with 1.5 L of arsine results in AsH$_x$ moieties (where x = 0 to 3) selectively adsorbing within the patterned area, producing a saturation arsenic density of 0.24 ML. An image of this surface is shown in **Figure 3**c. Following arsine dosing, the exposed region of the surface has a different appearance, consisting of a high density of features, whose identity is hard to distinguish due to their tight and disordered packing. An enlargement of the upper right corner of the patterned and dosed area is shown in **Figure 3**e. It is clear from these images that arsenic containing molecules from the gas phase have attached to the silicon surface where the hydrogen resist has been removed, while the resist protects the surface from AsH$_3$ adsorption elsewhere.

In the final step, shown in **Figure 3**d, annealing the surface to 350°C for 2 minutes results in substitutional incorporation of arsenic and the ejection of silicon onto the surface, exclusively within the patterned area. Once again, an enlargement of the upper right-hand corner of the patterned and dosed area (this time post-anneal) is shown in **Figure 3**d. Two examples of short dimer chains consisting of ejected silicon atoms are highlighted with arrows. Confirmation of the successful execution of this final step demonstrates that an arsenic dopant layer can be substitutionally incorporated into an arbitrarily defined region of a silicon surface, where the shape of the region is defined by the hydrogen desorption pattern written by an STM tip.

## 5. Encapsulation of Arsenic Layers: Low Temperature Silicon Epitaxy

The final, *in situ*, ultra-high vacuum step in the STM device fabrication is encapsulation with crystalline silicon. Once the arsenic atoms are incorporated into the silicon surface through a patterned hydrogen resist, the dopant structure must then be buried beneath a few nanometers of epitaxial silicon. This protective capping layer is grown via molecular beam epitaxy, using a solid intrinsic silicon sublimation source (SUSI-40, MBE Komponenten GmbH).

The thermal budget of the sample during the epitaxy stage is tightly restricted by two competing requirements: 1) dopant segregation and diffusion must be minimized so that the substitutional arsenic atoms remain in position – demanding a minimization of sample temperature; 2) crystallographic defect formation must be minimized to ensure that arsenic donor atoms remain electrically activated – usually achieved by elevated temperature overgrowth.[51] In the case of phosphorous, these two requirements can be reasonably satisfied by holding the sample temperature constant at 250 °C, throughout. In the case of arsenic we find (via secondary ion mass spectrometry; SIMS) that this approach is inadequate. We attribute this to the fact that arsenic is considerably more prone to surface segregation.[52] As described below, we have determined a variable sample temperature program during overgrowth that suppresses arsenic surface segregation during silicon epitaxy, allowing for the growth of *both* well confined *and* electrically activated arsenic δ-layers.

To achieve thermal control of arsenic segregation and diffusion during silicon epitaxy, a number of saturation arsenic delta layers were overgrown with 15 nm of silicon at a rate of 1 ML/min, using different sample temperature programs. Arsenic concentration depth profiles were subsequently measured using SIMS, and **Figure 4** shows SIMS data for four different temperature programs as described by the temperature vs time insets in each SIMS profile. **Figure 4**a shows the result of silicon overgrowth at a constant sample temperature of 240 °C. This arsenic distribution is dominated by a high and narrow peak located at a depth of ~1 nm. Beneath this narrow surface peak the arsenic concentration reduces to a much lower, but not an insignificant constant level (> 1 × $10^{18}$ atoms/cc) before terminating in a slight peak at the δ-layer origin depth, 15nm below the surface. Beyond this depth, the arsenic concentration rapidly drops below the instrument's detection limit (1×$10^{18}$ atoms/$cm^3$). From this we conclude that the arsenic layer is not effectively buried beneath the silicon, but rather, a large percentage of the 0.24 ML of arsenic atoms undergo continual cycles of surface segregation, i.e. they exchange positions with newly arrived silicon adatoms, and assume energetically preferred positions within the top layer.[53,54]

**Figure 4**b shows the arsenic profile for overgrowth at a constant but lower temperature of 190 °C. Here we observe a shift in weight from the surface to the subsurface arsenic peak, and find that the arsenic concentration is no longer uniform in the region between peaks. Below the surface peak, it slowly rises to an asymmetric peak at the arsenic origin depth of ~15nm, before rapidly falling off below the detection limit. Segregation is an activated process, and by lowering the constant growth temperature the segregation is found to be partially suppressed.

This arsenic segregation suppression can be further enhanced through a greater reduction in substrate temperature during encapsulation. **Figure 4**c shows the arsenic profile resulting from overgrowth at the lowest achievable sample temperature given our silicon source – sample geometry. In this case, there is no Joule heating of the sample, only the unavoidable radiative heating from the silicon MBE source. The substrate begins overgrowth at 40 °C increasing to a stable 100 °C after the first 12 minutes of growth (12 ML). From the SIMS profile, we see that arsenic surface segregation is highly suppressed in this sample, resulting in a very well confined (<2 nm FWHM) and symmetric peak (~ 5 × $10^{20}$ atoms/cc) located at the origin depth of 15 nm. This peak, which falls rapidly to below the instrument's detection limit within 3 nm in either direction, represents the *best achievable confinement* of the arsenic δ-layer grown in our MBE system. However, due to the reduced substrate temperature during overgrowth, and the high concentration of arsenic surface atoms, the critical thickness at which the epitaxial layer becomes amorphous will also be reduced considerably.[55] At a temperature of 100 °C, and deposition rate of 1 ML/min, the critical thickness is expected to reduce below 2 nm,[51] meaning the upper 13 nm or more of silicon will be amorphous, and unsuitable for device encapsulation.

In order to facilitate suppression of arsenic segregation during reduced temperature silicon overgrowth, yet still achieve epitaxy and avoid amorphization, we have developed the strategy (adapted from a similar concept used for phosphorus δ-layers in silicon [56,57,58]) of growing a very thin, reduced temperature "locking layer" that is kept below the critical amorphization thickness. With a reduced temperature locking layer in place, the substrate is then heated to a temperature that can support epitaxial growth for the remainder of the overlayer deposition. **Figure 4**d shows an example of locking layer assisted epitaxial growth applied to an arsenic δ-layer. Here the sample is not actively heated for the first 10 minutes (10 ML for a 1 ML/min growth rate) after which it is then rapidly heated to 250 °C, and held constant for the remaining 230 minutes of overgrowth. Comparing the SIMS profiles of **Figure 4**c and d, we see that increasing the temperature to 250 °C following growth of a 10 ML locking layer, has no detrimental effect on the confinement. The locking layer strongly suppresses arsenic surface segregation, producing epitaxially capped arsenic δ-layers with very tight confinements; confinements which are limited primarily by the solid-state diffusivity of arsenic in silicon. It is interesting to note here that arsenic has a lower diffusion coefficient than phosphorus in bulk silicon[23], which suggests that with the segregation issue overcome, using arsenic in place of phosphorus should allow for fabrication of denser and more tightly confined dopant structures, in every dimension.

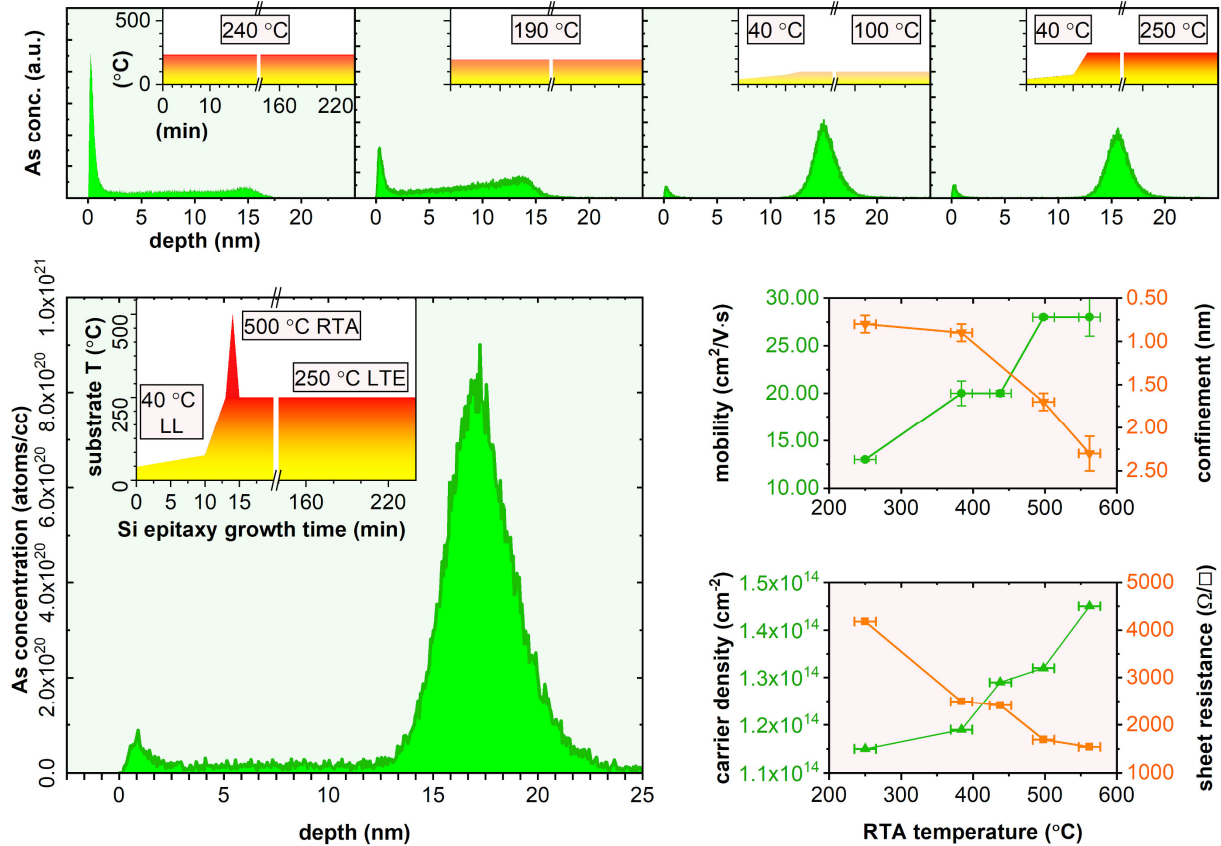

*Figure 4: Arsenic δ-layer distribution and electronic transport controlled through a silicon overgrowth thermal annealing program*

*a-e) SIMS profiles of five different saturation arsenic δ-layer samples overgrown with 15 nm Si, at a deposition rate of 1 ML/min. Sample temperature program during silicon epitaxy was modified for each sample according to the inset temperature vs time profiles. f) δ-layer mobility and confinement as a function of RTA temperature. g) δ-layer carrier density and sheet resistance as a function of RTA temperature. Error bars in the plots in panels f and g account for temperature measurement uncertainty and magnetoresistance curve fitting confidence levels as explained in the supplementary material, section b.*

## 6. Electrical Activation and Contacting of Arsenic δ-layers

To establish full compatibility of arsine and arsenic with the STM device fabrication process, the buried arsenic dopant structures must be electrically activated and Ohmic contacts achieved with the arsenic layer beneath the silicon cap. Satisfaction of this requirement has been verified by measurement of linear and symmetric IV curves during standard magnetotransport characterization of the arsenic δ-layers. As explained in detail in the supplementary material, Hall bars were fabricated from locking layer encapsulated arsenic δ-layers, and Ohmic contacts achieved through deposition of aluminum directly onto the δ-layer edge.

In addition to fabricating Ohmic contacts, we have also optimized the transport properties of the arsenic δ-layers for improved device performance. The optimizations were achieved by improving the epitaxial silicon quality and minimizing crystallographic defects, resulting in a maximization of carrier density and mobility, and a minimization of sheet resistance. Despite the fact that the locking layer thickness used is below the critical amorphization limit, including this reduced temperature step in the overgrowth is detrimental to crystal quality, and thus to the electrical performance of the δ-layers. To overcome this issue, we include a rapid thermal anneal (RTA) stage in the overgrowth sample temperature program, as has been done for phosphorus δ-layers.[56,57,58]

**Figure 4**e shows the SIMS profile of an arsenic δ-layer grown using both locking layer and RTA stages in the substrate temperature program. The temperature vs time profile used here is shown in the inset of **Figure 4**e. It is similar to the profile shown in the inset of **Figure 4**d, with the exception of the additional 500 °C × 10 second RTA, occurring between the locking layer growth and constant temperature epitaxy stages. Comparing the SIMS profiles of **Figure 4** (d) and (e) we find that introduction of this RTA stage is not detrimental to the enhanced confinement provided by the locking layer in any significant way. The effect of the RTA on transport properties is however significant.

**Figure 4**f and g show the effect of the RTA temperature maximum on δ-layer carrier concentration, sheet resistance, mobility, and confinement. The data for these figures of merit were obtained from Hall measurements on five arsenic δ-layers, with heating profiles identical to that of **Figure 4**e, except with each using a different RTA peak value, covering the range from 250 °C to 560 °C. For a RTA greater than 500 °C, the electronic layer thickness increases beyond 2 nm, while the improvement in sheet resistance and mobility has levelled out. For this reason, the RTA temperature maximum is considered optimum at ~500 °C. Despite the fact that both the mobility and sheet resistance values plateau at $1.6 \pm 0.1$ kΩ/□, and $28 \pm 1$ cm$^2$/(V·s) respectively, the carrier density continues increasing beyond this optimum RTA temperature. At the highest RTA temperature maximum examined, 560 °C, we measure a carrier density of $1.45 \times 10^{14}$ cm$^{-2}$, which corresponds to an arsenic donor atom density of only 0.21 ML. Given our estimate of arsenic saturation coverage at 0.24 ML, this maximum value represents an activation of 88% of the donor atoms. At the optimized RTA temperature of 500 °C, this activation percentage is even lower at 81%. Nevertheless, this carrier concentration is well above the metal insulator transition, and greater than ½ of the typical carrier density of $2.4 \times 10^{14}$ cm$^{-2}$ found in phosphorus δ-layers. The fact that 100% of the donor atoms in the arsenic δ-layers is not surprising, as arsenic is known to suffer from clustering at high concentrations,[59,60] resulting in incomplete activation at concentrations $> 10^{20}$ cm$^{-3}$ (or $2 \times 10^{13}$ cm$^{-2}$) in traditional implantation or diffusion doping of silicon.[51]

Given the large parameter space involved in optimizing arsenic δ-layer growth for both confinement and electrical transport, there is potentially room for improvement in performance values beyond those presented here. However, comparing the figures of merit achieved for arsenic δ-layers with those typically achieved for phosphorus δ-layers,[56,58] we find that they are at most a factor of 2 worse, and that arsine and arsenic are 100% compatible with atomic-scale STM device fabrication, and therefore represent a viable alternative to phosphine and phosphorous.

CONCLUSIONS

We have demonstrated the full compatibility of arsine as a precursor gas for the atomic-scale positioning of arsenic donors in silicon using STM-based hydrogen-desorption lithography, and, when combined with low temperature silicon epitaxy, the capability to fabricate buried, atomic-scale, dopant structures in silicon. Room temperature dissociation of arsine on Si(001) differs from the dissociation of phosphine on Si(001) in that full dissociation of arsine occurs much more rapidly at room temperature. This has interesting implications, including the potential for tighter control of atomic-scale placement of arsenic compared to phosphorus. We find also that saturation arsine coverage is self-limiting at 24% ML, invariant with dose rate, and that monolayer hydrogen termination acts as an effective resist layer for the adsorption of arsine on Si(001). We show that arsenic can be incorporated in the silicon surface lattice by annealing at 350 °C. We demonstrate successful low temperature silicon epitaxial overgrowth by growing arsenic δ-layers, where segregation was suppressed through the use of an unheated, ultra-thin, locking layer. A rapid thermal anneal stage of 500 C × 10 sec immediately following locking layer growth improves electronic transport properties resulting in 2 nm thick, conductive layers with sheet resistance, and carrier concentration of 1.5 kΩ, and $1.3 \times 10^{14}$ cm$^{-3}$, respectively. It has been a decade and a half since phosphorus dopant placement using STM H-lithography was first demonstrated and the work presented here shows that using arsenic as the n-type dopant in place of phosphorus is not only possible, but has demonstrable advantages over phosphorus. Finally, the parameters used for arsenic and for phosphorus device fabrication are completely compatible with one another, demonstrating the possibility of multispecies dopant devices.

EXPERIMENTAL METHODS

**Sample Preparation and Scanning Tunneling Microscopy:** Si(001) samples were diced to 2×9 mm from a 0.3 mm thick, Czochralski grown, bulk phosphorus doped wafer with a resistivity of 1-10 Ω·cm. Samples were cleaned ultrasonically *ex situ* in acetone followed by isopropyl alcohol, then thermally outgassed *in situ* (base pressure < $1 \times 10^{10}$ mbar) for >8 hours at 600 °C, and finally flash annealed multiple times at 1200 °C, using direct current resistive sample heating. Sample temperature was monitored using an infrared pyrometer (IMPAC IGA50-LO plus), with a total estimated measurement uncertainty of +/-30 °C. Silicon epitaxy was performed at a base pressure of $1 \times 10^{10}$ mbar, using an all silicon, solid sublimation source (SUSI-40, MBE Komponenten GmbH) operated at a deposition rate of 1.5 Å/min. During silicon overgrowth, sample temperature was indirectly monitoring by measuring sample resistance, while the sample was heated using direct current resistive sample heating. All scanning tunneling microscopy measurements were performed in an Omicron variable temperature series STM at room temperature with a base pressure of $<5 \times 10^{11}$ mbar.

**Density Functional Theory Calculations:** All DFT calculations reported here were conducted using the Gaussian 09 software [48] and methods of energy computation and geometry optimization implemented therein. We calculate adsorption energies for various arsine configurations on the silicon (001) surface using an approach that we refer to as the cluster composite model (CCM).[13,61] In this approach, the Si(001) surface is represented using hydrogen-truncated cluster models, such as $Si_{15}H_{16}$ or $Si_{54}H_{44}$, and various correction terms are applied to minimize the effects of finite cluster size. We note that the specific approach used here for the $AsH_3$/Si(001) system is the same as in our previous work on the $PH_3$/Si(001) system,[13] and the reader is referred to this article for the full technical details. Reaction steps (i) in Fig. 2g was calculated using what is referred to as the 3D2R cluster model in Ref.[13] while reaction steps ii), iii), and iv) used the 2D1R cluster model with the hemihydride dimer represented by a separate cluster (see Eq. 5 in Ref. [13]).

**Kinetic Monte Carlo Simulations.** KMC methods are explained in detail in the supplementary material.

**Secondary Ion Mass Spectrometry:** Time of flight (ToF) SIMS measurements were made using an IONTOF ToF.SIMS[5] system with a 25 keV $Bi^+$ primary ion beam in high current bunch mode (HCBM), and a 500 eV, 35 nA $Cs^+$ sputter beam. Depth profiles were made with a 300 µm$^2$ sputter crater, and the analytical region was gated to the central 50 µm$^2$ of the sputter region

**Hall bar Fabrication and Magnetotransport Measurements:** Arsenic delta layer samples were fabricated into 6 terminal Hall bars using standard UV photolithography, with electrical contact made via aluminum thermally evaporated over the edge of the mesa structures. Magneto-transport measurements were made at a temperature of 5 K in a Cryogenics Ltd. cryogen free measurement system using magnetic fields up to 5 Tesla. Details of these measurements, and quantification of the delta layer electrical transport characteristics are explained in detail in the supplementary material.

ASSOCIATED CONTENT

**Supporting Information**: Kinetic Monte Carlo calculations and magneto-transport data analysis are each explained in detail in the supporting information document.

AUTHOR INFORMATION

The authors declare no competing financial interests.

ACKNOWLEDGMENT

This projected has been supported by the EPSRC project Atomically Deterministic Doping and Readout For Semiconductor Solotronics (grant number EP/M009564/1). P.C.C. and E.V.S.H. were partly supported by the EPSRC Centre for Doctoral Training in Advanced Characterisation of Materials (grant number EP/L015277/1), and also by Paul Scherrer Institute and IHP – Leibniz-Institut für innovative Mikroelektronik, respectively. O.W. acknowledges the support of the Australian Research Council (ARC) Centre of Excellence for Quantum Computation and Communication Technology (project number CE110001027). Computing support was provided by the Australian National Computational Infrastructure (NCI). All data created during this research is openly available at zenodo.org (DOI: 10.5281/zenodo.3406575).

# Supporting Information:

## a) Kinetic Monte Carlo Simulations

Computer simulations of the evolution of an $AsH_3$-exposed Si(001) surface on time and length scales relevant to the STM experiments were conducted using the kinetic Monte Carlo (KMC) method[1] and a discrete lattice representation of the adsorbed molecular species. Key elements of this model are illustrated in Supplementary Figure S1.

The lattice used to represent the adsorbed species is shown in Figure S1 (a) with a unit cell (gray background shading) that corresponds to the (2x1) reconstruction of the Si(001) surface, and much larger super cells used in our simulations to represent an extended surface. Square boxes in Figure S1 (a) indicate discrete *lattice sites* that represent the surface binding sites for an $AsH_3$ molecule or any of its fragments. These sites come in three types, namely, *dimer-end*, *end-bridge*, and *dimer-bridge*, labeled 'a', 'b', and 'c' within the unit cell in Figure S1(a), respectively. Each of the sites exists in one of seven *states*, which indicates the type of molecular species occupying the site. The states used here are defined in Figure S1(b) and they represent $AsH_3$, $AsH_2$, $AsH$, $As$, and $H$ molecular species, as well as a state F indicating a free or unoccupied site. Further defined is a distinct state X, which serves as a blocking state to indicate that a site is unavailable for occupation by another species. This is used in our model to block dimer-end sites when an adjacent dimer-bridge or end-bridge site is occupied, as such a bridging adsorbate would utilize the dangling bond valences at the two dimer ends involved in the bridge.

Together, the lattice and the configuration of states at all sites provides a compact representation of an extended Si(001) surface with a coverage of $AsH_3$ molecules in varying degrees of dissociation. An example configuration is shown in Fig. S1(c), describing a low-coverage surface with two $AsH_3$-related species, namely, a molecular $AsH_3$ molecular adsorbate at a dimer-end position near the top of the figure and a fully dissociated end-bridge As+3H species near the bottom.

In a KMC simulation, a given initial configuration of states is advanced in time using a set of discrete *transformations* that occur with a defined rate[1]. In our case, the initial configuration is the bare surface without any adsorbates (i.e. all sites in the F state) and the set of transformations corresponds to a set of discrete chemical reactions that may occur upon $AsH_3$ exposure, broadly divided into adsorption, dissociation, and rearrangement reactions. These transformations are set out in Figure S1(d) to (o), and they each comprise a pattern of specific states at a subset of lattice sites, a prescription for how these states change when the reaction occurs, and a parameter to define the rate of reaction. Note that an empty box in a transformation pattern indicates that the site can be in any state for the pattern to match.

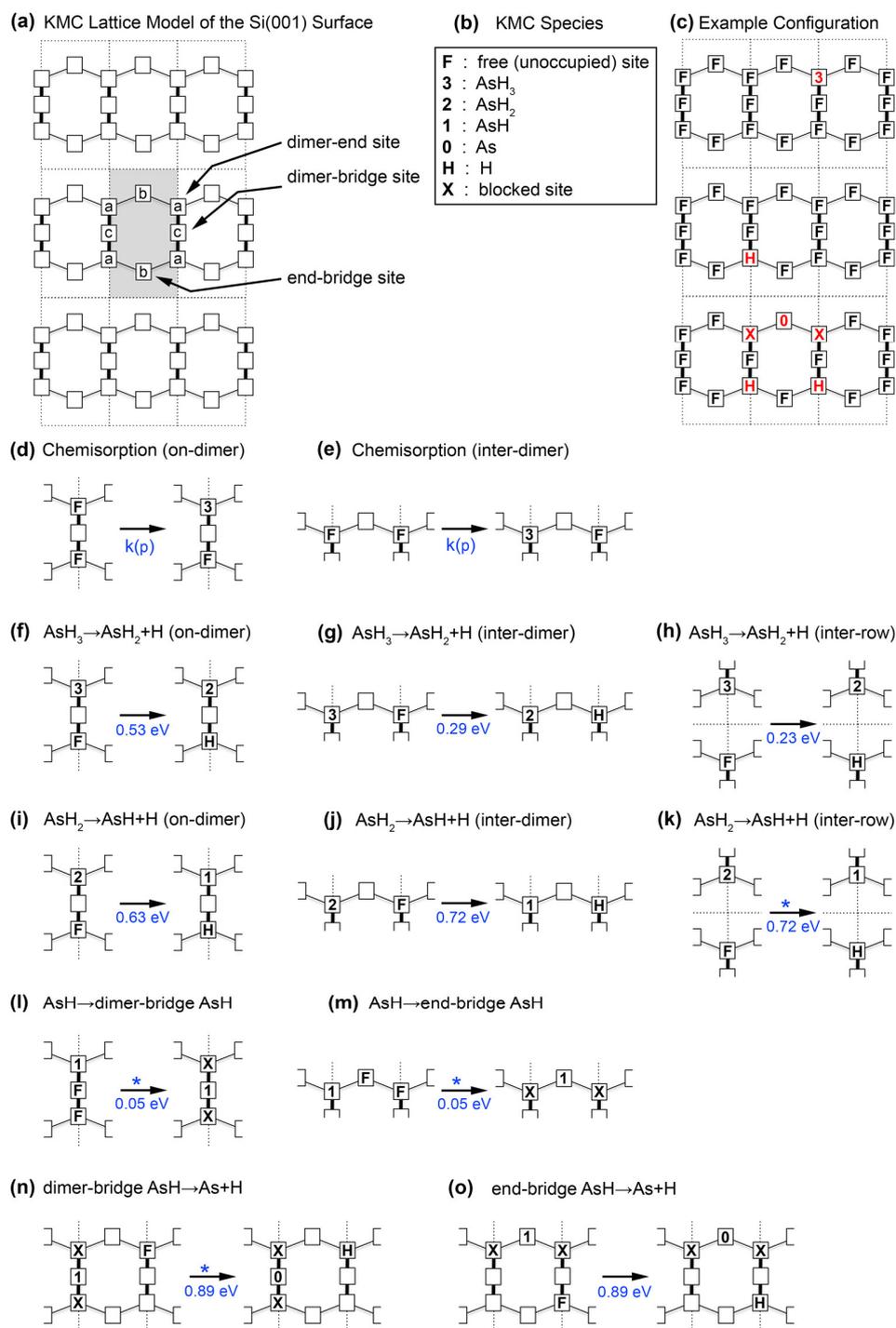

***Figure S1: KMC model of the Si(001) surface exposed to AsH3.*** *(a) The lattice representation of binding sites (indicated by squares) on the Si(001) surface with the unit cell shaded gray. (b) The set of seven possible occupation states at each binding site. (c) An example configuration in a 3x3 supercell featuring a molecular AsH₃ adsorbate and the end-bridge As+3H structure observed as the type-1 STM feature. (d,e) KMC transformations representing two types of chemisorption processes with the rate of transformation given by the pressure-dependent surface impingement rate. (f-o) KMC transformation representing various hydrogen-shift and rearrangement reactions relevant to AsH₃ dissociation on Si(001) with the rate of reaction defined using Arrhenius activation energies.*

For all other transformations [Figure S1 (f) to (o)] the rates are defined by an activation energy, which is converted into a rate constant using the Arrhenius equation with an assumed attempt frequency of $10^{13}$ s$^{-1}$ and a temperature of 298 K. The activation energies for most of the transformations are derived from the density functional theory calculations of suitable prototype reactions that represent the transformation. Here, these prototype reactions are those calculated for the main AsH$_3$ dissociation pathway (cf. main text Fig. 1g) and some of the competing reaction pathways discussed in the main text. For other transformation which are indicated by an asterisk ('*') in Fig. S1, the activation energies are derived by analogy to a similar reaction within the set. For example, the activation energy of the dimer-bridge AsH→As+H transformation [Fig. S1(n)] assumed to be the same as that calculated by DFT for the end-bridge AsH→As+H transformation [Fig. S1(n)].

One of the key tasks of a KMC simulation is to assemble a list of all possible transformations for a given configuration. This is accomplished by identifying all those transformation patterns that match the given configuration. Using again the example configuration in Fig. S1(c), the only transformation patterns that apply are the two types of adsorption reactions [Fig. S1(d) and (e)] matching many F-state sites in this configuration as well as the dissociation reactions [Fig. S1(f) and (g)] for the molecular AsH$_3$ adsorbate near the top of Fig. S1(c). None of the transformation patterns match to the end-bridge As+3H structure near the bottom of the figure, which reflects the fact that this structure represents an end point in the dissociation pathway of a single AsH$_3$ molecule.

From the assembled list of all applicable transformations, a single transformation is selected at random and executed to advance the system to the next configuration, and the time is advanced also. A new list of applicable transformations is then assembled for the new configuration and the process repeats until the desired simulation time has been reached, or the system has reached a stable point where none of the transformations apply. Critically, both the random selection of the transformation and the time advancement are performed such that the all transformations occur at the specified rate when examined using in a sufficiently large ensemble average (see (Gillespie, 1976) for details). The sequence of configurations that result from successive transformations applied in the course of a KMC simulation create a single trajectory in which AsH$_3$ molecules adsorb and dissociate in a stochastic manner. The pressure-dependent species coverages reported in the main text Fig. 2c using large 40x40 super cells and an average over 100 trajectories for each pressure, which sufficiently removes stochastic effects from the data.

All KMC simulations were carried out using a software developed by one of the authors (O.W.).

## b) Hall and Magnetoresistance Measurements of As δ-layers

Eight-terminal Hall bars were fabricated from multiple Si:As delta layer samples and electrically characterized by IV, Hall, and magnetoresistance measurements. The Hall bar geometry is illustrated below in Figure S22, alongside an example of a two-terminal IV measurement (terminals 8-4, sample d) demonstrating Ohmic contacting to the As delta layer.

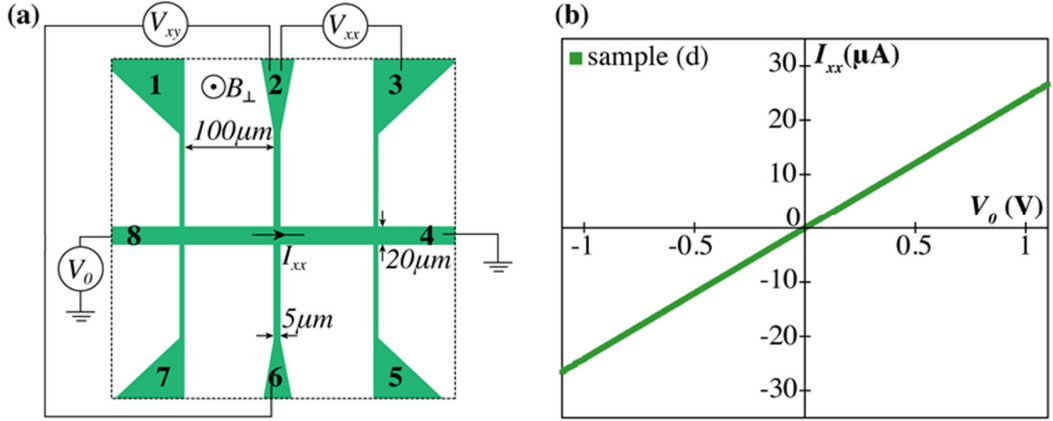

*Figure S2: Eight-terminal Hall bar geometry and two-terminal IV measurement on a Si:As delta layer. a) The orientation of the perpendicular magnetic field $B_\perp$, and the measurement terminals for current and voltages $I_{xx}$, $V_{xx}$, and $V_{xy}$ are indicated on the Hall bar schematic. b) A two-terminal IV measurement (across terminals 8 and 4) on sample d (RTA = 500 °C) measured at 5 K, is both linear and symmetric indicating that the delta layer contacting is Ohmic.*

Figure S3 shows Hall and magnetoresistance measurements for all six As delta-layer Hall bar samples along with fits to the data. The approach used in fitting to the weak-localisation data follows the methods of Sullivan et al.,[2] and assumes that the signal depends on both the magnitude and direction of the applied magnetic field, **B**, where the change in the conductance, $\delta\sigma_{xx}$, is for a weak spin-orbit donor. From the Hall and magnetoresistance measurements, $\Delta\sigma_{xx}$ is obtained from the measured resistivity $\rho_{xx}$:

$$\Delta\sigma_{xx}(\mathbf{B}) = \frac{1}{\rho_{xx}(\mathbf{B})} - \frac{1}{\rho_{xx}(0)}. \tag{1}$$

For a perpendicular magnetic field, the change in conductance is:

$$\delta\sigma(B_\perp) = \left(\frac{e^2}{2\pi^2\hbar}\right)\left[\Psi\left(\frac{1}{2} + \frac{\hbar}{4eB_\perp L_\phi^2}\right) - \Psi\left(\frac{1}{2} + \frac{\hbar}{2eB_\perp L^2}\right) + \ln\left(\frac{2L_\phi^2}{L^2}\right)\right], \tag{2}$$

where $\Psi$ is the digamma function, $L_\phi$ the dephasing length and $L$ the mean free path. For a parallel magnetic field, the change in conductance is:

$$\delta\sigma(B_\parallel) = \left(\frac{e^2}{2\pi^2\hbar}\right)\ln(1 + \gamma B_\parallel^2), \tag{3}$$

where $\gamma$ is the fitting parameter.

Once the fits to the experimental weak-localisation data are determined, the fit parameters can be used to obtain an approximate measure of the δ-layer thickness:

$$T = \left(\frac{1}{4\pi}\right)^{1/4}\left[\left(\frac{\hbar}{eL_\phi}\right)^2\left(\frac{L}{L_c}\right)\gamma\right]^{1/2}, \quad (4)$$

where $L_c$ is the correlation length, assumed to be of the order of the mean donor spacing $L_c = 1/\sqrt{n}$, where $n$ is the carrier concentration extracted from the experimental Hall-measurements shown in the insets of Figure S3.

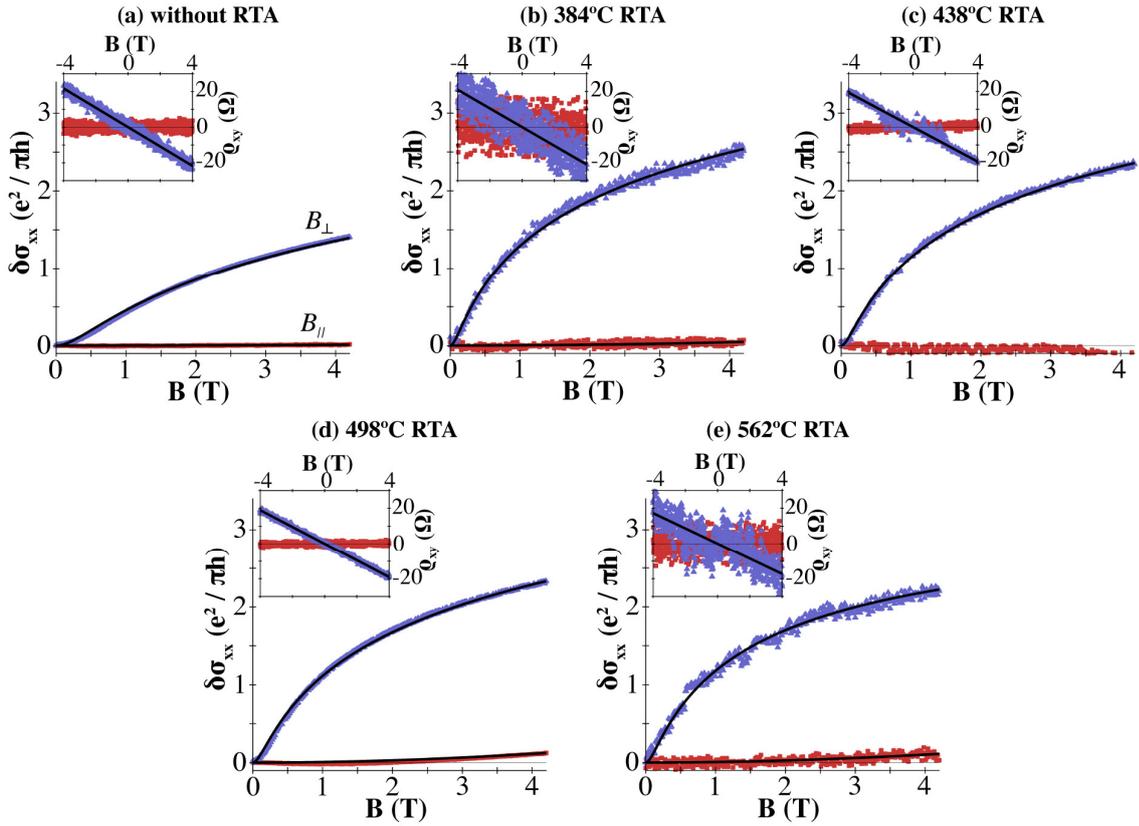

*Figure S3: Summary of weak-localization data for parallel- (red) and perpendicular- (blue) magnetic fields.* (a)-(d) The primary figures show the weak-localization data, whereas the insets show the Hall-measurements, with progressively increasing rapid thermal anneal (RTA) temperatures during the silicon overgrowth. The fits are applied from the equations provided in (Sullivan et al., 2004). No fit could be made to the parallel field data in (c).

Table S1 summarises values extracted from the Hall measurements ($n$, $\mu$, and $R_0$), and the best fit parameters for the weak-localization magnetoresistance plots for each of the arsenic δ-layers grown using different rapid-thermal-anneal (RTA) temperatures. Using equations (1)-(3), a least-squares method is used to fit to the experimental weak-localisation data and the uncertainties in the fit parameters reflect the confidence intervals dictated by the fit parameters. For samples (a)-(b) and (d)-(e), there is a small feature present in the magnetoresistance near zero-field regime, which yields a negative change in conductance. Establishing the source of this behaviour in the Si:As system is beyond the scope of the current work and is the focus of ongoing investigations. For the purposes of the present work, all fits to extract thickness estimates were made by omitting the near zero-field behaviour, which is reflected in the uncertainties of the fit parameters shown in Table S1. For sample (c), no fits could be made to extract thickness estimates to the parallel field as the change in conductance is always negative.

| Sample RTA (°C) | Electron density ($n$) ($10^{14}$/cm$^2$) | Sheet Resistance ($R_0$) ($10^3$ Ω/□) | Mobility ($\mu$) (cm$^2$/V·s) | Mean free path ($L$) (nm) | Dephasing length ($L_\phi$) (nm) | Fitting Parameter ($\gamma$) ($10^{-4}$ T$^{-2}$) | Mean Thickness ($T$) (nm) |
|---|---|---|---|---|---|---|---|
| (a) 250±15 | 1.15 ± 0.02 | 4.18 ± 0.02 | 13.0 ± 0.2 | 7.6 ± 0.2 | 30.9 ± 0.1 | 6.5 ± 0.4 | 0.8 ± 0.1 |
| (b) 384±15 | 1.19 ± 0.07 | 2.49 ± 0.04 | 20.0 ± 1.3 | 8.1 ± 0.7 | 58.6 ± 0.6 | 27.0 ± 3.5 | 0.9 ± 0.1 |
| (c) 438±15 | 1.29 ± 0.02 | 2.41 ± 0.02 | 20.0 ± 0.3 | — | — | — | — |
| (d) 498±15 | 1.32 ± 0.01 | 1.69 ± 0.02 | 28.1 ± 0.2 | 7.1 ± 0.3 | 52.2 ± 0.2 | 75.0 ± 1.0 | 1.7 ± 0.1 |
| (e) 562±15 | 1.45 ± 0.10 | 1.54 ± 0.06 | 28.0 ± 2.0 | 16.0 ± 0.7 | 55.3 ± 0.8 | 67.0 ± 7.3 | 2.3 ± 0.2 |

*Table S1: As delta-layer electrical transport characteristics extracted from the magnetoresistance data shown in Figure S3. The electron density (n) and mobility (μ) are determined from the fits to the inset data in Figure S3. The sheet resistance (R_0), mean-free path (L), dephasing length (L$_\phi$) and parallel field parameter (γ) are determined from the experimental fits to the primary data shown in Figure S3. The mean thickness (T) is then determined as in (Sullivan et al, 2004).*

## Supporting Information References